\documentclass[11pt,a4paper]{JHEP3}

\usepackage{amsmath,amssymb}

\newcommand{\be}{\begin{equation}}
\newcommand{\ee}{\end{equation}}

\newcommand{\bea}{\begin{eqnarray}}
\newcommand{\eea}{\end{eqnarray}}

\def\({\mbox{$(\!($}}
\def\){\mbox{$)\!)$}}

\title{ Selfduality of non-linear electrodynamics with derivative corrections}
\author{ W. A. Chemissany, Joost de Jong, and Mees de Roo   \\ 
Centre for Theoretical Physics, University of Groningen,\\
Nijenborgh 4, 9747 AG Groningen, The Netherlands\\
 E-mail: \email{W.Chemissany@rug.nl, JoostdeJong@gmail.com, M.de.Roo@rug.nl }}

\preprint{
\hepth{0610060}\\
UG-06/07}
 

\abstract{In this paper we investigate how
electromagnetic duality survives derivative corrections
to classical non-linear electrodynamics. In particular, 
we establish that electromagnetic selfduality is satisfied 
to all orders in $\alpha'$ for the four-point function 
sector of the four dimensional open string effective 
action.}

\begin{document}

\section{Introduction\label{Intro}}

The symmetry between electric and magnetic fields is a 
fundamental property of Maxwell theory, and of certain 
extensions such as the nonlinear electrodynamics of
Born and Infeld \cite{Born}.
Consider the free Maxwell equations in
$d=4$ flat space:
\footnote{Throughout the paper, we are working in
d=4, and we adopt the following convention: Minkowski metric with
$diag(-,+,+,+)$ signature, where $\widetilde{\widetilde{F}}=-F$,
and often we use the notation ${\rm tr}\,FG=-F_{ab}G^{ab}$.}
\begin{equation}
\label{INT1}
\partial_{a}F^{ab}=0\,\qquad\partial_{a}\tilde{F}^{ab}=0\,,
\end{equation}
where $F_{ab}=\partial_{a}A_{b}-\partial_{b}A_{a}$ and 
$\widetilde{F}$ is the Hodge dual of $F$, i.e,
$\tilde{F}_{ab}=\frac{1}{2}\epsilon_{abcd}F^{cd}$.
Indeed, (\ref{INT1}) is invariant under the Hodge duality
transformation. 

One can pose the question whether a generalization of duality
invariance continues to hold for deformations of the Maxwell action.
In particular, one might consider Lagrangian densities depending
only on abelian field and a deformation parameter,
say $\alpha'$, which coincides with the Maxwell Lagrangian
for $\alpha'=0$.   
The general Lagrangian satisfying such restrictions
and leading to electromagnetic duality invariance involves an
arbitrary real function of one real argument
\cite{Gibbons,Zumino,Kuzenko}. A particular
example is the Born-Infeld Lagrangian\ \cite{Born}:
\begin{equation}\label{INT2}
\mathcal{L}=1-\sqrt{-\text{det}(\eta_{ab}+ \alpha'\,F_{ab})}\,.
\end{equation} 
The duality invariance of Born-Infeld theory was established 
in \cite{Schrodinger}.  

Our purpose in this paper is to extend the deformations of
the Maxwell theory to also include derivatives of the field strength
$F$, and to investigate if the duality invariance can be preserved.
This is relevant for the application in string theory, where it is
known that the open string string effective action, which for 
slowly varying
fields coincides with the Born-Infeld action, also contains
derivative corrections. We establish in this paper that the
effective action for the open string 4-point function, truncated to
four dimensions, satisfies the property of electromagnetic duality
also when derivative corrections are included.

In Section \ref{EMSD} we will briefly review basic definitions and results
concerning electromagnetic duality. The duality of the terms related to the
4-point function will be discussed in Section \ref{FPF}. In Section
\ref{Conc} we discuss the extension to higher-order contributions to the
string effective action.

\section{Electromagnetic Selfduality\label{EMSD}}

In this section, we  briefly review some definitions and results,
see also
\cite{Gibbons,Zumino,Kuzenko,Gaillard} and references
therein. We consider the Lagrangian to be a function $\mathcal
{L}$ of one variable $F$.
The field equation and Bianchi identity are
\begin{equation}
\label{paper1}
\partial_{a}G^{ab}=0\,,\qquad
\partial_{a}\tilde{F}^{ab}=0\,,
\end{equation}
where $G$ is an
anti-symmetric tensor of rank two defined by 
\begin{equation}
\label{paper2}
G^{ab}=-\frac{\partial\mathcal{L}(F)}{\partial F_{ab}}\,.
\end{equation}  
Consider transformations which
send the pair $(G,F)$ to $(G',F')$: 
\begin{equation}
\label{paper5}
\begin{pmatrix}G'(F')\\\tilde{F}'\end{pmatrix}
=D\begin{pmatrix}G(F)\\\tilde{F}\end{pmatrix}\,,
\end{equation}
with $D\in \mathop{\rm
GL}(2,\mathbb{R})$.  
We can solve (\ref{paper5}) for $(G,F)$ in terms of $(G',F')$ and 
then substitute in equations (\ref{paper1})-(\ref{paper2}) to find
the transformed version of the field equations and Lagrangian. 
We will
assume the existence of the transformed Lagrangian ${\cal L}'(F')$,
which satisfies
\begin{equation}\label{paper22}
G'(F')^{ab}=-\,\frac{\partial\mathcal{L'}(F')}{\partial F'_{ab}}\,.
\end{equation}
Then the property of electromagnetic duality invariance or
selfduality is defined by
\begin{equation}
   \mathcal{L'}(F)=\mathcal{L}(F)\,.
\label{selfdual}
\end{equation}
From (\ref{selfdual}) and the  form of the duality transformations
one derives that the
duality symmetry is $SO(2)$, that the constraint
\begin{equation}
\label{constraint} 
\text{tr}\,G\tilde{G}= \text{tr}\,F\tilde{F}
\end{equation}
must hold, and that the combination
\begin{equation}
\label{invariant}
\mathcal{L}(F)-\frac{1}{4}\text{tr}\,F G
\end{equation}
is duality invariant.

It is natural to ask what happens  if the action also 
depends on derivatives of the field strength.  At first sight, 
it seems
that the analysis presented is not applicable anymore, and
hence should be modified. This also happens in the extension
which includes additional scalars, e.g, axion and dilaton fields
\cite{Gibbons,Rasheed}. In that case the duality symmetry is
modified and becomes $SL(2,\mathbb{R})$.
In the case of derivative corrections 
however, most of the discussion above can be taken over if one 
works with the action rather than with
the Lagrangian density, and uses functional differentiation 
\cite{Kuzenko}.
We then define\footnote{We use 
the chain rule and \begin{equation}\label{p3}
\frac{\delta}{\delta F_{ab}(x)}F_{cd}(y)=2\delta_{cd}^{ab}\,\delta(y-x)\,.
\end{equation}}
\begin{equation}\label{p5}G^{ab}=
  -\frac{\delta}{\delta F^{ab}} S[F]\,.
\end{equation}
The duality transformations retain the same form, 
we now assume that $G'$ follows from an action $S'[F']$, 
the duality condition then reads
\begin{equation}\label{p6} 
S'[F]=S[F]\,,
\end{equation}
while the constraint (\ref{constraint}) becomes:
\begin{equation}\label{p7}
\int d^4 x\,\text{tr}\,G\tilde{G}=\int d^4 x \,\text{tr}\,F\tilde{F}\,.
\end{equation}
In Appendix \ref{SDcondition} we outline a proof of (\ref{p7}), and discuss the 
effect of field redefinitions. If we have an action $S_0$ satisfying the condition
of selfduality, then of course any action related to that action
by a field redefinition should also be considered to be
electromagnetically selfdual. In Appendix A we show that this implies that 
we should allow (\ref{p7}) to hold up to terms containing
$\partial_a G_0{}^{ab}$, the equation of motion of $S_0$. 

\section{Selfduality of the 4-point function derivative corrections
\label{FPF}}

In this Section we will extend the electromagnetic
selfduality of the open superstring effective action to include 
derivative corrections. The terms we will consider have the generic 
form
\begin{eqnarray}\label{pp1}
\mathcal{L}_{(m,n)}&=&\alpha'^{m}\,\partial^{n}F^{p}\,,\qquad 
\text{for}\,\,\,p=m+2-n/2\,,
\\
\mathcal{L}_{m}&=&\alpha'^{m}F^{m+2},\qquad \text{for}\,\,\, n=0\,.
\end{eqnarray}
The absence of corrections with $n=2$ has been established in
\cite{AndTs}. All corrections for $n=4$ have been constructed by
Wyllard \cite{Wyll}, while it is known that terms with $p$ odd
are absent. The terms with $p=4$ have been obtained to all orders in
$\alpha'$ \cite{dRE}. We will establish in this section the electromagnetic
duality of the $p=4$ terms. 

Of course electromagnetic selfduality as discussed in this paper holds in
spacetime dimension $d=4$, while the superstring corrections are obtained
in $d=10$. We will discuss electromagnetic duality of the contributions
of the form (\ref{pp1}), setting all other ten-dimensional fields to zero,
and truncating the result to $d=4$, i.e., by restricting
the Lorentz index values to $d=4$. Furthermore, the result can hold only
order-by-order in $\alpha'$, in the sense that for each order of $\alpha'$
the corresponding $p=4$ contribution to the effective action satisfies, together
with the $m=0$ Maxwell term, electromagnetic selfduality to order $m$ in
$\alpha'$.

Let us start the discussion with the $m=4$ terms. Then the four-derivative terms 
are
\begin{equation}
\label{p17}
\mathcal{L}_{(4,4)}=a_{(4,4)}\alpha'^{4}
t_8{}^{abcdefgh}
\partial_{k}F_{ab}\partial^{k}F_{cd}\partial^{l}F_{ef}\partial_{l}F_{gh}\,,
\end{equation}
where $t_8{}^{abcdefgh}$ is antisymmetric in the pairs $ab$, $cd$,
etc., and is symmetric under the exchange of such pairs. 
It expands as follows for arbitrary antisymmetric matrices $M_i$:
\begin{eqnarray}
\label{p19} 
&&t_8{}^{abcdefgh}
M_{1\,}{}_{ab}M_{2\,}{}_{cd}M_{3\,}{}_{ef}M_{4\,}{}_{gh}=8\, 
(\text{tr}M_{1}M_{2}M_{3}M_{4}+
\text{tr}M_{1}M_{3}M_{2}M_{4}+
\text{tr}M_{1}M_{3}M_{4}M_{2})
\nonumber
\\&&\qquad
-2\,(\text{tr}M_{1}M_{2}\text{tr}M_{3}M_{4}+
\text{tr}M_{1}M_{3}\text{tr}M_{2}M_{4}+
\text{tr}M_{1}M_{4}\text{tr}M_{2}M_{3})\,.
\end{eqnarray}
The values of the constants $a_{m,2m-4}$ can be found in 
\cite{dRE}. 
The equation of motion of the combination $\mathcal{L}_0+\mathcal{L}_{(4,4)}$
contains
\begin{equation}
\label{pp18}
  G^{ab} = F^{ab}+ G_{(4,4)}^{ab}\,,\quad
  G_{(4,4)}^{ab}=
  4 a_{(4,4)}\alpha'^{4}t_8{}^{\,abcdefgh}
  \partial_{k}\,(\partial^{k}F_{cd}\partial^{l}F_{ef}\partial_{l}F_{gh})\,.
\end{equation}
To establish electromagnetic selfduality
we have to establish that (\ref{p7}) holds. It only makes sense
to verify this to order $\alpha'^{4}$, 
since in higher orders other
contributions to the effective action would interfere.
Since the $\alpha'^{0}$ terms in (\ref{p7}) cancel we have to 
verify that
\begin{equation}
\label{p118}
I=\int d^{4}x\, \text{tr}\,\widetilde{F}G_{(4,4)}=0\,.
\end{equation}
After partial integration (\ref{p118}) takes on the form 
\begin{equation}
\label{p218}
I=\int d^{4}x \,t_8{}^{\,abcdefgh}
\partial_{k}\widetilde{F}_{ab}\partial^{k}F_{cd}
\partial_{l}F_{ef}\partial^{l}F_{gh}=0.
\end{equation}

The crucial property, which in fact holds to all orders in 
$\alpha'$, is that in $\mathcal{L}_{(m,2m-4)}$ the 
indices of the fieldstrengths $F$ are all contracted amongst 
each other, and therefore also the derivatives are contracted 
\cite{dRE}. The identity
($F_{k}\equiv\partial_{k}F$)
\begin{equation}
\label{p22}
(\widetilde{F}_{k}{F}_{l}+\widetilde{F}_{l}{F}_{k})_{a}\,^{b}=
-\tfrac{1}{2}\delta_{a}^{b}\,\text{tr}\,\widetilde{F}_{k}{F}_{l}\,,
\end{equation}
in combination with the complete symmetry of $t_8$, can then 
be used to express the traces over four matrices resulting from the expansion of 
(\ref{p218}) in terms of products of traces over two matrices. 
This leads to the required cancellation.

For higher orders in $\alpha'$ the $p=4$ terms contain more derivatives, 
but again these are all contracted with each other, while the
tensor structure of the fieldstrengths remains the same. Essentially
one has to show that 
\begin{equation}
\label{p23}
t_8{}^{\,abcdefgh}\Big[
\widetilde{F}_{1\,}{}_{ab}F_{2\,}{}_{cd}F_{3\,}{}_{ef}F_{4\,}{}_{gh}+
F_{1\,}{}_{ab}\widetilde{F}_{\,2}{}_{cd}F_{\,3}{}_{ef}F_{4\,}{}_{gh}+
F_{1\,}{}_{ab}F_{2\,}{}_{cd}\widetilde{F}_{3\,}{}_{ef}F_{4\,}{}_{gh}+
F_{1\,}{}_{ab}F_{2\,}{}_{cd}F_{3\,}{}_{ef}\widetilde{F}_{4\,}{}_{gh}\Big]\,,
\end{equation} 
where the subscripts $1,2,3,4$ indicate the derivative structure,
vanishes. Using again (\ref{p22})
and the symmetry of $t_8$ one establishes that (\ref{p23})
vanishes independently of the precise way the derivatives are
contracted.

This gives the desired result: electromagnetic duality survives,
to this order in $\alpha'$, the addition of derivative corrections.
Note that in verifying (\ref{p7}) to order 
$\alpha'^4$ there are no term proportional to 
$\partial_aG_0{}^{ab}$ left over. Had we started from the
$p=4$ terms in a different basis, for instance that given in
\cite{Wyll} for $m=4$, then indeed (\ref{p7}) would hold only
up to terms that vanish on-shell.

\section{Conclusions\label{Conc}}

It would be of interest to use electromagnetic selfduality 
to constrain, or
to determine, the derivative corrections to the Born-Infeld action that 
are not known explicitly. However, it is well-known that already the
Born-Infeld action itself is not the only selfdual deformation of the 
Maxwell action, the ambiguity can be parametrized by a real function of 
one variable \cite{Gibbons}. From the previous section it clear
that $\mathcal{L}_{(m,2m-4)}$ is not the only $p=4$ action
with derivative corrections that satisfies (\ref{p7}) to order
$\alpha'^4$. Indeed, we found that the result depends only on the
presence of the tensor $t_8$ and on that fact that there are no contractions 
between derivatives and fieldstrengths. The result is independent
of the precise way the derivatives are placed.

Given these ambiguities, it is clear that electromagnetic duality
can only constrain but not determine the derivative corrections
to the terms related to the six-point function, $p=6$. For the
four-derivative terms $n=4$ we do have the result of \cite{Wyll}.
The method used in Section \ref{FPF} is however not applicable,
because the the 
property of having no contractions between field strengths
and derivatives no longer holds. Nevertheless, it would be interesting
to extend the analysis of selfduality to those terms. 

Another extension would be to add derivative corrections to the
$SL(2,\mathbb{R})$ invariant extension of Born-Infeld \cite{Rasheed}.
This problem is currently under investigation \cite{Chem}.

\acknowledgments
This work supported by the European
Commission FP6 program MRTN-CT-2004-005104 in which the authors
are associated to Utrecht University.

\appendix

\section{Integral Form of the selfduality condition
and field redefinitions\label{SDcondition}}

We derive the consistency condition (\ref{p7}) for
infinitesimal duality transformations. 
Then 
\begin{equation}
\label{p8} G'^{ab}[F']=G^{ab}[F]+\lambda \tilde{F}^{ab}\,,\quad
\tilde{F}'^{ab}=\tilde{F}^{ab}-\lambda G^{ab}[F] \,,
\end{equation}
 with
\begin{equation}
\label{p9}
G'^{ab}[F']=-\frac{\delta S'[F']}{\delta F'_{ab}(x)}\,.
\end{equation}
Selfduality (\ref{p6}) implies
\begin{align}
\label{p10}
G'^{ab}[F',x]=-\frac{\delta S[F']}{\delta F'_{ab}(x)}
= -\Big(\frac{\delta S[F]}{\delta F'_{ab}(x)}+\frac{\delta}{\delta
F_{ab}(x)}\delta S[F]\Big)\,,
\end{align} 
where we use
\begin{equation}
\label{p11}
\delta S[F]=S[F']-S[F]
\end{equation}
and
\begin{equation}
\label{p12}
\frac{\delta}{\delta F'_{ab}(x)}\delta S[F]= 
\frac{\delta}{\delta F_{ab}(x)}\delta S[F]+
\mathcal{O}(\lambda^2)\,.
\end{equation}
$\delta S[F]/\delta F'_{ab}$ can be evaluated 
as follows:
\begin{eqnarray}
\label{p13}
\frac{\delta S[F]}{\delta F'_{ab}(x)}&=&
\tfrac{1}{2}\int d^4 y
\frac{\delta S[F]}{\delta F_{cd}(y)}
\frac{\delta F_{cd}(y)}{\delta F'_{ab}(x)}
\nonumber\\
&=&
\tfrac{1}{2}\int d^4 y 
\frac{\delta S[F]}{\delta F_{cd}(y)}\frac{\delta}{\delta F'_{ab}(x)}
\Big(F'_{cd}(y)-\lambda \tilde{G}_{cd}[F,y]\Big)
\nonumber\\
&=& -G^{ab}[F,x]+
\tfrac{\lambda}{2}\int d^4 y \,G^{cd}[F,y]
\frac{\delta}{F_{ab}(x)}\tilde{G}_{cd}[F,y]
\nonumber\\
&=&-G^{ab}[F,x]+
\tfrac{\lambda}{4}\frac{\delta}{F_{ab}(x)} \Big
(\int d^4 y G^{cd}[F,y]\tilde{G}_{cd}[F,y]\Big).
\end{eqnarray}
Substituting (\ref{p13}) in (\ref{p10}) yields
\begin{equation}
\label{p14} 
G'^{ab}[F',x]=G^{ab}[F,x]-\frac{\delta}{F_{ab}(x)}\Big(\delta S[F]+
\tfrac{\lambda}{4}\int d^4 y
G^{cd}[F,y]\tilde{G}_{cd}[F,y]\Big).
\end{equation} 
On the other
hand, from the variation (\ref{p8}) of $G$ it follows
\begin{equation}
\label{p15} 
G'^{ab}[F',x]=G^{ab}[F,x]-\frac{\delta}{F_{ab}(x)}
\Big(-\tfrac{\lambda}{4} \int d^4 y
F^{cd}(y)\tilde{F}_{cd}(y)\Big).
\end{equation} 
The variation of $S$ is
\begin{equation}
\label{p16}
\delta S[F]=\tfrac{1}{2}\int d^4 y 
\frac{\delta S[F]}{\delta F_{cd}(y)}\delta F_{cd}(y)=
-\tfrac{\lambda}{2}\int d^4 y
\,G^{cd}[F,y]\tilde{G}_{cd}[F,y]
\end{equation} 
Inserting
(\ref{p16}) into (\ref{p14}) and comparing the resulting
expression to (\ref{p15}), one finds the integrated form
selfduality condition (\ref{p7}).

If we have a selfdual action $S_0$ satisfying (\ref{p7}),
and we perform a field redefinition on the vector potential,
(\ref{p7}) will only be satisfied modulo terms proportional
to $\partial_a G_0{}_{ab}$. To see this, write the new action
as
\begin{equation}
  S=S_0+S_1\,,
\end{equation}
where $S_1$ is of the form
\begin{equation}
  S_1=\int d^4x\,V_b[F,x]\partial_aG_0{}^{ab}\,.
\end{equation}
Then 
\begin{equation}
  G^{ab}(x) = G_0{}^{ab}(x)
  - \int d^4y\,\left(\frac{\delta V_d(y)}{\delta F_{ab}(x)}\,
               \partial_cG_0{}^{cd}(y)
         - \partial_cV_d(y)
           \frac{\delta G_0{}^{cd}(y)}{\delta F_{ab}(x)}\right)\,.
\end{equation}
Using the fact that $S_0$ satisfies (\ref{p7}) the remaining
condition for the selfduality of $S$ is
\begin{equation}
 0 = \int d^4xd^4y\,\left(
 \widetilde{G}_0{}_{ab}(x)\frac{\delta V_d(y)}{\delta F_{ab}(x)}\,
               \partial_cG_0{}^{cd}(y)
   - \partial_cV_d(y)
         \frac{\delta G_0{}^{cd}(y)}{\delta F_{ab}(x)}
          \widetilde{G}_0{}_{ab}(x)
                     \right)\,.
\label{app7}
\end{equation}
The second term in (\ref{app7}) vanishes. 
This can be seen by using
\begin{equation}
   \frac{\delta G_0{}^{cd}(y)}{\delta F_{ab}(x)}=
   \frac{\delta G_0{}^{ab}(x)}{\delta F_{cd}(y)}
\end{equation}
and (\ref{p7}) for $S_0$, the result then contains 
$\partial_a\widetilde{F}^{ab}$ which vanishes. The remaining term
is proportional to $\partial_cG_0{}^{cd}(y)$ so that indeed we
see that selfduality holds modulo the $S_0$ equation of motion.

\end{document}